
%
%
%
%
\input epsf.tex
\input harvmac.tex
%
\def\Titlemz#1#2{\nopagenumbers\abstractfont\hsize=\hstitle\rightline{#1}%
\vskip 0.4in\centerline{\titlefont #2}\abstractfont\vskip 0.3in\pageno=0}
%
\Titlemz{\vbox{\baselineskip 12pt \hbox{LAVAL-PHY-22/94}}}
{\vbox {\centerline{ Integrability test for spin chains} }}
\centerline{M. P. Grabowski and P. Mathieu$^*$}
\smallskip
\centerline{ \it D\'epartement de
Physique, Universit\'e Laval, Qu\'ebec, Canada G1K 7P4}
\vskip.4in
\bigskip
\centerline{\bf Abstract}
\bigskip\vskip 0.45cm
We examine a simple heuristic test of integrability for
quantum chains.  This test is applied to a variety of systems,
including a generic isotropic spin-1 model with nearest-neighbor
interactions and a multiparameter
family of spin-1/2 models generalizing  the XYZ chain,
with next-to-nearest neighbor interactions
and bond alternation. Within the latter family
we determine all the integrable models with an $o(2)$ symmetry.

\bigskip\vskip 0.5cm

{\noindent PACS numbers: 75.10.Jm, 11.10.Lm, 11.30.-j. }
\vskip 2.2 in
{\noindent $^*$ Work supported by NSERC (Canada).  }
\noindent\smallskip
\Date{11/94}

\vfill\eject

\newcount\eqnum \eqnum=1
\def\eq{
\eqno(\secsym\the\meqno)
\global\advance\meqno by1
 }
\def\eqlabel#1{
{\xdef#1{\secsym\the\meqno}}
\eq
}

\newwrite\refs
\def\startreferences{
 \immediate\openout\refs=references
 \immediate\write\refs{\baselineskip=14pt \parindent=16pt \parskip=2pt}
}
\startreferences

\refno=0
\def\aref#1{\global\advance\refno by1
 \immediate\write\refs{\noexpand\item{\the\refno.}#1\hfil\par}}
\def\ref#1{\aref{#1}\the\refno}
\def\refname#1{\xdef#1{\the\refno}}

\def\s{\sigma}

\def\la{\lambda}

\def\ra{\rightarrow}

\let\n=\noindent
\def\frac#1#2{{\textstyle{#1\over #2}}}

\font\smallcapfont=cmr9
\def\sc#1{{\smallcapfont\uppercase{#1}}}

\def\Tr{{\rm Tr}}

\def\text#1{\quad\hbox{#1}\quad}


\def\ubrackfill#1{$\mathsurround=0pt
	\kern2.5pt\vrule depth#1\leaders\hrule\hfill\vrule depth#1\kern2.5pt$}
\def\contract#1{\mathop{\vbox{\ialign{##\crcr\noalign{\kern3pt}
	\ubrackfill{4pt}\crcr\noalign{\kern3pt\nointerlineskip}
	$\hfil\displaystyle{#1}\hfil$\crcr}}}\limits
}

\def\ubrack#1{$\mathsurround=0pt
	\vrule depth#1\leaders\hrule\hfill\vrule depth#1$}
\def\dbrack#1{$\mathsurround=0pt
	\vrule height#1\leaders\hrule\hfill\vrule height#1$}
\def\ucontract#1#2{\mathop{\vbox{\ialign{##\crcr\noalign{\kern 4pt}
	\ubrack{#2}\crcr\noalign{\kern 4pt\nointerlineskip}
	$\hskip #1\relax$\crcr}}}\limits
}
\def\dcontract#1#2{\mathop{\vbox{\ialign{##\crcr
	$\hskip #1\relax$\crcr\noalign{\kern0pt}
	\dbrack{#2}\crcr\noalign{\kern0pt\nointerlineskip}
	}}}\limits
}

\def\ucont#1#2#3{^{\kern-#3\ucontract{#1}{#2}\kern #3\kern-#1}}
\def\dcont#1#2#3{_{\kern-#3\dcontract{#1}{#2}\kern #3\kern-#1}}

\def \sumL{{\sum_{j\in\Lambda}}}
\def \S{{\bf {\sigma} }}
\def \a{{\la_x}}
\def \al{{\alpha}}
\def \b{{\la_y}}
\def \c {{\la_z}}

\font\tenmib=cmmib10
\font\sevenmib=cmmib10 at 7pt
\font\fivemib=cmmib10 at 5pt
\newfam\mibfam 

\textfont\mibfam=\tenmib
\scriptfont\mibfam=\sevenmib
\scriptscriptfont\mibfam=\fivemib
\mathchardef\alphaB="080B
\mathchardef\betaB="080C
\mathchardef\gammaB="080D
\mathchardef\deltaB="080E
\mathchardef\epsilonB="080F
\mathchardef\zetaB="0810
\mathchardef\etaB="0811
\mathchardef\thetaB="0812
\mathchardef\iotaB="0813
\mathchardef\kappaB="0814
\mathchardef\lambdaB="0815
\mathchardef\muB="0816
\mathchardef\nuB="0817
\mathchardef\xiB="0818
\mathchardef\piB="0819
\mathchardef\rhoB="081A
\mathchardef\sigmaB="081B
\mathchardef\tauB="081C
\mathchardef\upsilonB="081D
\mathchardef\phiB="081E
\mathchardef\chiB="081F
\mathchardef\psiB="0820
\mathchardef\omegaB="0821
\mathchardef\varepsilonB="0822
\mathchardef\varthetaB="0823
\mathchardef\varpiB="0824
\mathchardef\varrhoB="0825
\mathchardef\varsigmaB="0826
\mathchardef\varphiB="0827

\def \calS{{\cal {S}}}
\def \calO{{\cal {O}}}

\def \hal{{\frac {1} {2}}}

\def \pr{{^\prime}}
\def \bis{\pr\pr}

\def \r#1{{}}

\def \a{{\la_x}}
\def \b{{\la_y}}
\def \c {{\la_z}}
\def \S{\sigmaB }
\def \sb{\sigmaB}

\vfill\eject

\newsec{Introduction}

This work examines integrability criteria for quantum chains.
In particular, we propose a simple heuristic test of integrability for
quantum chains with short-range interactions, consisting
essentially in proving or disproving the existence of
a nontrivial conserved local  charge involving three-point interactions.
We indicate the conditions in which this might be a
necessary condition  of integrability, as well as
circumstances in which it is expected to be a sufficient condition.
The test is then applied to a number of models.
Our approach is motivated by our recent extensive
analysis of  the structure of the conservation laws for
integrable spin chains of the XYZ type and the Hubbard model
[\ref{Grabowski M P and  Mathieu P 1994
{\it Mod. Phys. Lett.} {\bf 9A} 2197;
{\it The structure of the conservation laws in integrable spin chains with
short-range interactions} preprint Laval-PHY-21/94
(hep-th 9411045)}\refname\GMa].

For hamiltonian systems, the common
definition of quantum integrability
mimics  the Liouville-Arnold definition of classical
integrable hamiltonian systems:
a quantum system with $N$ degrees of freedom is called integrable, if it
possesses $N$ nontrivial, functionally independent and mutually
commuting conservation laws.

For classical continuous systems,
the proof of integrability usually amounts to
displaying a Lax or zero-curvature formulation.
Although this is not always easy,
there are various other manifestations of integrability that can
be probed, such as a
bi-hamiltonian formulation, nontrivial symmetries,
prolongation structures  or
higher order conserved charges
(see e.g. [\ref{ Zakharov V E 1981 {\it What is integrability}
(Springer Verlag);
Fordy A P (ed) 1990 {\it  Soliton Theory: a Survey of Results}
(Manchester University Press) part VI}]).  Furthermore, one can apply a
systematical integrability
test, based on the Painlev\'e property
[\ref{ Ablowitz M J,  Ramani A and Segur H 1980
{\it J. Math. Phys.} {\bf 21} (1980) 715; Weiss J, Tabor M and Carnevalle G
1983 {\it J. Math. Phys.} {\bf 24} 522;  Newell A C , Tabor M and  Zeng Y B
1987 {Physica } {\bf D 29} 1}].
For classical discrete systems, there are related integrability
indicators
[\ref{Ramani A, Grammaticos B and
Tamizhmani K M 1992 {\it J. Phys. A:  Math. Gen.} {\bf 25} L883}].
But except for higher order conservation laws,
these integrability signals are no longer available for quantum chains.

On the other hand, the
integrability of quantum chains is usually demonstrated rather
indirectly, by
showing that the model can be solved by the coordinate Bethe ansatz
or that the
hamiltonian can be derived from a commuting family of transfer
matrices related to the
Yang-Baxter equation.  But these  are only sufficient
conditions for integrability.  Moreover,
testing these sufficient conditions is often not
easy.\foot{Note however, that the Yang-Baxter equation
implies a relatively simple
equation, known as the Reshetikhin condition.
See section 2 for a discussion of this point.}

It is thus clearly desirable to design a more general, simple and efficient
integrability test for quantum chains.  Here we propose a simple test based
on the existence of  a local nontrivial three-point charge $H_3$, the higher
order conservation law just
above $H_2$, the defining hamiltonian of the model.  Locality
means that the interaction involving a certain set of sites disappears when the
distances between them become sufficiently large.
We will indicate below why for quantum chains, one can expect that the
existence of $H_3$ should be
a necessary and in some cases a sufficient
condition for integrability.
 From the computational point of view,
the advantage of this simple-minded approach is that the number of possible
candidates for the three-point charge is usually not exceedingly large,
and they are often restricted by the symmetries of the system.

A related integrability test has been studied in
[\ref{Fuchssteiner B and Falck U  1988 in {\it Symmetries and
Nonlinear Phenomena} ed D Levi and P Winternitz (World Scientific)
}\refname\FF].
 However, these authors confined their test
to the search of one nontrivial conservation law (not necessarily $H_3$)
in spin 1/2 models,
 with
the simultaneous existence of a ladder operator providing a recursive scheme
for
 the
calculation of the other conservation laws.  This is certainly less general and
less constructive than the test proposed here.

Another integrability test has been formulated
for a very special class of self-dual quantum chains.
For such systems, there exists a simple
sufficient condition for integrability,
due to Dolan and Grady [\ref{ Dolan L and Grady M 1982 {\it Phys. Rev.} {\bf
25D} 1587}].  This condition actually applies to any
type of self-dual systems, discrete or continuous,
and defined in any number of space-time dimensions.
These systems are described by an
hamiltonian of the form:
$${H_2=\alpha H+\beta \tilde {H};}\eqlabel\sdham$$
where $\tilde H$ is the dual of $H$, with duality being defined as
any nontrivial linear operator with the
property $\tilde{ \tilde H}=H$.
If such hamiltonians satisfy the relations:
$${ [H, [H, [H, \tilde H]] ]=16 \; [H, \tilde H],}\eqlabel\sdeq$$
there exists a (potentially infinite, for infinite systems)
 family of conservation laws and these can be constructed systematically.
We stress that this sufficient condition for integrability
is simply the  requirement  of
the existence of a charge $H_3$ of a particular form.
We have thus a neat situation here in which the
the existence of a third-order charge $H_3$ guarantees
the existence of an infinite family of conserved charges.
Unfortunately, the applicability of this condition seems rather limited.\foot
{The condition (\sdeq) appears to apply only to the $Z_n$ generalization of the
Ising model
defined in [\ref{von Gehlen G and Rittenberg V 1985 {\it Nucl. Phys.}
{\bf B257} 351}].
Systems satisfying (\sdeq) are sometimes called superintegrable and the
underlying
algebraic structure is the so-called Onsager algebra
[\ref{McCoy B M 1993 in
{\it Important Developments in Soliton Theory} ed. A S Fokas and V E Zakharov
(Springer Verlag) p. 418}].
}
 For example, the
$H_{XYZ}$ hamiltonian is self-dual, as it may be put in the form:
$${H_{XYZ}=\alpha H_{YZ}+\beta \tilde H_{YZ};}\eq$$
where
$$H_{YZ}=\sumL [\b \s^y_j \s^y_{j+1} +\c \s^z_j \s^z_{j+1}] ,\eq$$
where $\s_i^a$ are Pauli spin matrices
 acting  nontrivially only on the site $i$ of the
lattice $\Lambda$.
The duality is defined by:
$${ \tilde \s^x_j=\s^y_j,\quad \tilde \s^y_j=\s^x_j,\quad
\tilde \s^z_j=-\s^z_j.}\eq$$
However, the condition (\sdeq) is not satisfied unless $\b=0$
(in which case (\sdham) reduces
to the Ising model hamiltonian), or $\c=0$ (the XY model).


Our proposal is described in detail in the next section.
It is formulated for the class of translationally
invariant models with nearest-neighbor interactions.
These models are
defined on a lattice
$\Lambda$ which is either finite with periodic boundary conditions
(i.e. $\Lambda=\{1,\dots,N\}$, with
$N+1\equiv 1$) or infinite (i.e. $\Lambda=Z$).
Notice that the restriction to nearest-neighbor interactions
is not as
severe as it might appear at first
sight, since this class of models actually contains,
when appropriately reformulated, any
model involving binary interactions with finite range.
Indeed, any model with a finite interaction range $k_0$ (where
$k_0=2$ corresponds to  nearest-neighbor interactions)  can be
equivalently described in terms of nearest-neighbor interactions
just by grouping together
$k_0$ consecutive sites into a single one on which a vectorial spin-like
 variable would live.
Similarly, any model with a more general invariance under
a shift $j\to j+j_0$  can be made invariant
with respect to a translation by a single unit,
by grouping together $j_0$ consecutive sites.

 Let $\{S_i^a\}$ denote
a set of quantum operators (distinguished by the index $a$),
acting nontrivialy only in a Hilbert space $V_i$.
 (We will assume in this work
that all the $V_i$'s  are finite-dimensional; in general they don't have to be
identical).
The full space of states of such chain is then the tensor
product $\bigotimes_{i\in\Lambda} V_j$.
The class of hamiltonians
under consideration  has then the following  form:
$$H_2=\sumL [g_j+h_{j,j+1}],\eqlabel\hamclass$$
where $g_j=g(S_j)$ describes interactions at site $i$
and $h_{j,j+1}$
is some site-independent function of $S_j, S_{j+1}$ which
describes nearest-neighbor interactions.


\newsec{Conjectured integrability tests for quantum chains}

\subsec{ A conjectured necessary condition for integrability of a
quantum chain}

{}From the early days of
soliton theory, it has been clear that the existence of a nontrivial
conservation law
(beyond those associated with the standard conservation of mass or charge,
momentum and
energy) is a very strong indication of
 the existence of an infinite number of additional
nontrivial conservation laws.  But for continuous systems, either classical or
quantum, one cannot
focus the attention on the existence or non-existence of the conservation law
with a
degree just above the one which usually plays the r\^ole of the hamiltonian.
%
%
The reason is simply that symmetry considerations can prevent
the existence of $H_n$ for a set of values of $n$.  Take for instance the KdV
equation,
$$ u_t+u_{xxx}+6 u u_x=0,\eq$$
for which $u$ has degree 2 in the normalization where $\partial_x$ has degree
1. Due to a
hidden
$Z_2$ invariance, there are no conservation laws with densities of odd degree.
%
%
  More generally, for Toda systems
related to Lie algebras, the values of $n$ for which $H_n$ is
non-zero are related to the exponents of the corresponding Lie algebra [\ref{
Drinfeld V G and  Sokolov V V 1985 {\it J. Sov. Math} {\bf 30} 1975;
 Wilson G  1981 {\it Ergo. Theor. Dynam. Syst.} {\bf 1} 361}].
But notice that when the
conservation laws are calculated by a recursive method, the conserved densities
associated
to charges that vanish by symmetry are not found to be zero; rather they are
total
derivatives.

As indicated above, these
considerations apply for both classical and quantum systems.  But in the
quantum case,
there is another possible source for the absence of a conserved charge of given
order,
which is that the density can be exactly proportional to a null vector. Again
this
is observed in the quantum KdV case, where for special values of the central
charge, the
free parameter in the defining commutation relation, some conserved densities
 become
proportional to  null vectors of the Virasoro algebra [\ref{Freund P G O,
 Klassen T R and
Melzer E, 1989 Phys. Lett. {\bf B229} 243; Eguchi T and  Yang S K,
1990 {\it Phys. Lett.} {\bf B235}
282; Kuniba A,  Nakanishi T and  Suzuki J 1991  {\it Nucl. Phys.}
 {\bf B356} 750;  Di
Francesco P and Mathieu P 1992 {\it Phys. Lett.} {\bf B278} 79}].

Now, our point here is that both these
 possibilities are
absent in the case of quantum chains.
%
%
In particular, it is clear that
on the lattice there is no  room for total derivatives.
Furthermore, null vectors appear not to be
relevant if the spaces $V_i$ are finite-dimensional.\foot{
Constraints imposed on the factor spaces $V_i$
(e.g. realized via  a projection onto some subspace of $V_i$) may
``project out" some charges of an integrable model - in other words
some of the conserved  charges evaluated in a restricted space of
states may conceivably vanish. Such constraints would thus
have an effect similar as null-vectors.
However,
situations like this can   be avoided by
considering the conservation laws in the full (unrestricted)
Hilbert space.}
%

 On the other hand, for
systems whose integrability can be traced back to commuting transfer matrices,
the conserved
charges are obtained
from the expansion of the logarithm of the transfer matrix
[\ref{L\"uscher M 1976 {\it Nucl. Phys.} {\bf
B117}
475}\refname\Lush] and $H_3$ is
never absent. This can be easily seen for fundamental systems
[\ref{Korepin V E,
Bogoliubov N M and A. G. Izergin A G  1993
{\it Quantum Inverse Scattering Method and
Correlation
Functions} (Cambridge University Press) }\refname\KBI],
characterized by
a Lax operator $L_{n}(\la)$ proportional to the model's
$R$ matrix $R_{n0}(\la)$, normalized so that
$R_{n0}(0)=P_{n0}$:
$$L_{n}(\la)=R_{n0}(\la).\eq$$
Here we use the standard notation in which $R_{ij}$  acts in the
tensor product space $V_i \otimes V_j$, where $V_i$ is the vector space
associated to the
site $i$. The index zero refers to the quantum space $V_0$ on which the
matrix entries of $L_i$
act.
$P_{n0}$ denotes the permutation operator.
(For simplicity all the vector spaces $V_i$
are assumed to be equivalent here).
The transfer matrix is defined as:
$$T(\la) = \Tr_{V_0} L_N(\la)...L_1(\la).\eq$$
For the fundamental models, a simple calculation shows that
 the second logarithmic derivative of
the transfer matrix  at $\la=0$
is proportional to:
$$  Q_3 \sim \sumL [h_{j,j+1}, h_{j+1,j+2}]\eqlabel\qthreefun$$
where $h_{i,i+1}$ is defined in (\hamclass).
(Here we follow the notation used in [\GMa]: $Q_n$ stands for the $(n-1)$-th
derivative of the logarithm of the transfer matrix and $H_n$ is obtained from
$Q_n$ by stripping off contributions of the lower order charges.)    If one
excludes a pathological case in which all adjacent links commute,\foot{For
example, such pathological situation arises for the
``chopped XXZ" model introduced in section 3.}
the three-point
charge defined above is nontrivial. The existence of a nontrivial third order
charge
can also be proven for a more general class of integrable models,
characterized by
the condition $$L_{n}(0)=P_{n0}.\eqlabel\Linit$$
 Note that such $L$ matrix is not necessarily itself a solution of the
 Yang-Baxter equation.
An example of a transfer matrix which
satisfies (\Linit), but which does not define a fundamental
model, is provided by the
transfer matrix of the Hubbard model found by Shastry
[\ref{Shastry B S 1986   {\it Phys. Rev. Lett.} {\bf 56} 1529;
1986 {\it Phys. Rev. Lett.} {\bf 56}  2453; 1988 J. Stat. Phys. {\bf
50} 57}\refname\Shas].
For models satisfying (\Linit), the second logarithmic derivative
of the transfer matrix becomes:
$$ \eqalign{Q_3 &\sim\sumL [h_{j,j+1}, h_{j+1,j+2}]-
\sumL h_{j,j+1} h_{j,j+1} \cr &+
\sumL T^{-1}(0) \Tr_{V_0} (P_{N0}\dots{ d^2L_{j}\over {d\la^2}}(0) \dots
P_{10}).}\eq$$
The last two terms involve only nearest-neighbor interactions,
while the first is a sum over triples of consecutive spins.
This term has the same form as (\qthreefun), and it is clearly
nontrivial except again for pathological systems with all
adjacent links commuting.


  The above considerations motivate the
following conjecture for the class of systems described by
(\hamclass):
%

\n {\bf Conjecture 1}: A translationally invariant
periodic quantum spin chain, with  a hamiltonian
$H_2$  involving at most
nearest-neighbor interactions, is integrable
only if there exists a nonvanishing local independent charge $H_3$,
being a sum of terms coupling spins at most three sites,
which commutes with $H_2$ for all chain sizes $N\geq 3$.



This conjecture implies then a simple test consisting in establishing the
existence of such $H_3$.
In particular,
one may conclude
that a spin chain is nonintegrable, by demonstrating the
non-existence of a nontrivial $H_3$.
On the other hand, if such a charge exists, the system is likely to
be integrable; but of course its integrability has to be proven independently.
Actually, it appears also that for a large class of systems
(and in particular for all the models considered in sections 3 and 4),
the mere existence of $H_3$ is enough to guarantee the existence of
an infinite family of conserved charges in involution.


\subsec{Clarifying comments related to the formulation of conjecture 1}

Some elements entering in the formulation of the conjecture deserve
clarification.

\n i) {\it Independence of the charges}

\n
The charge $H_3$ in the above conjecture should be independent of the
hamiltonian and of possible charges of lower order (such as, e.g.
the components of the total spin).
For general quantum integrable models, the issue of functional
independence can be quite complicated; it is indeed a major
difficulty in formulating a
general definition of quantum integrability
in a completely rigorous way.\foot{
Dropping  the requirement of functional independence does not
lead to a meaningful definition of integrability [\ref{Weigert S 1992
Physica D {\bf 56 } 192}].
In the context of spin chains this can be easily seen. Consider
an isotropic chain, for which the hamiltonian commutes with all
the components of the total spin. Arbitrary powers
of any of the spin components yield then a set of mutually commuting
conserved charges. Removing the requirement of functional
independence in the definition of integrability
would therefore render any
isotropic spin chain automatically
integrable.}
However,
functional independence of the charges can be usually
easily verified for spin chains with short range interactions:
the leading term of $H_n$ contains $n$ adjacent
interacting spins; the leading terms of $H_n$ and $H_m$ for $m\not=n$ being
clearly distinct, the corresponding charges are linearly
independent.  Furthermore, such a
cluster of $n$ adjacent spins cannot be obtained from a
product of lower order charges
(since that would also generate terms with $n$ non-adjacent spins).


\n ii) {\it Stability of the charge under a variation of the chain lenght}

\n The last item in the above  conjecture ensures
that the existence of
$H_3$ should not be affected by a change $N\ra N+k$, where $k$ is an
arbitrary positive integer.
This stability requirement can be simply
illustrated for the XXX model defined by
the hamiltonian
$$H= \sum_{i=1}^N \sb_i\cdot\sb_{i+1}\eqlabel\Hxxx$$
(where as usual periodic boundary conditions are assumed). This hamiltonian
commutes with any component of the total spin
$$\calS^a= \sum_{i=1}^N \s_i^a.\eq$$
A simple calculation shows that for  $N=4$
the quantity
$$H'\equiv \sum_{i=1}^N \sb_i\cdot\sb_{i+2}\eqlabel\Hpxxx$$
is conserved.
$H'$ is also conserved for $N=5$ but not for any higher value of $N$.
The reason for this behavior is that
for $N=4$ and 5, $H'$ can be regarded as a nonlocal charge.
Such nonlocal charges typically can be obtained from powers of
the local ones. More exactly, for $N=4$ we have the following
identity (modulo an additive constant):
$$ \calS^a\calS^a = 2H+H'.\eq$$
Similarly, for $N=5$
$$ \calS^a\calS^a = 2H+2H'.\eqlabel\fiveid$$
However, for $N>5$ $H'$ is not related to $\calS^a\calS^a$ or other
nonlocal charges and it is no longer conserved.
For $N=6$ for instance, we have instead
$$ \calS^a\calS^a = 2H+2H'',\eq$$
where
$$H''\equiv \sum_{i=1}^N [\sb_i\cdot\sb_{i+2}+\hal \sb_i\cdot\sb_{i+3}]\eq$$
contains contributions with one and two holes.
Thus the conservation of $H'$ is not preserved
under the change $N\ra N+2$; the same is true for $H''$.
This is actually typical for all such ``accidental" charges, whose
conservation for some particular values of $N$ is due to
an accidental identity (relating them to nonlocal charges),
which is true only for these
particular values of $N$.

\n iii) {\it Locality vs nonlocality}

\n The above example illustrate another issue: it is not always easy to
distinguish the nonlocal
charges (e.g. $H'$) from the local ones (e.g. $H$)
for finite chains.
For an infinite chain,
local and nonlocal expressions are quite distinct (the latter contain
interactions between arbitrarily distant spins).
On the other hand, for
the XXX model with $N=4,5$, one hole in the expression of a 2-spin
conserved law reflects ``nonlocality"!
However, as exemplified above,
the form of those nonlocal charges that
can be written as powers of local ones is strongly $N$-dependent; this property
makes them easily detected. Notice also that
the local charges for a finite chain may be defined non-ambiguously
from the densities of the first $N$ charges of the infinite chain.


\subsec{A conjectured sufficient condition for integrability
based on the boost operator}

The integrable points in a multi-parameter space of  general spin chain
hamiltonians
can usually be simply characterized by the occurrence of a dynamical
symmetry.
Related to such symmetry is the existence of a ladder operator $B$,
acting on the conservation laws as
$$[B, H_n]=
H_{n+1},\eqlabel\heredi$$
where $H_n$ denotes a charge with at most $n$ adjacent interacting
spins.\foot{Note that
we allow  for the possibility
of a linear combination
of lower order charges $H_{m\le n}$ on the rhs of (\heredi).}
This
motivates the formulation of a simple conjectured sufficient condition for
integrability, based on the existence of a ladder
operator $B$, in conjunction with the
presence of a nontrivial charge $H_3$:

\n {\bf Conjecture 2}: A translationally invariant periodic
quantum chain,
with a hamiltonian $H_2$
involving at most nearest-neighbor interactions,
 is integrable if there exists an operator $B$ such that
for all chains of length $N\ge 3$,
$[B, H_2]$ is nontrivial and
$$[[B,H_2],H_2] =  0. \eq$$

In contrast with the first conjecture, the approach here is more constructive:
 it indicates how $H_3$ can be built, i.e.  as $[B,H_2]$.
This constructive aspect presupposes that we can easily guess the form
of $B$.
In all the cases we have considered, such a $B$
turns out to be proportional
to the first moment of an appropriately symmetrized form of
the hamiltonian $H_2$, i.e. with $$H_2 =\sumL h_{j,j+1},
\quad {\rm for} ~h_{j,j+1}~ {\rm symmetric ~ in}
{}~ j,~j+1,\eq$$
$B$ is found to be
$$ B=\sumL j~h_{j,j+1}.\eq$$
The condition of commutativity of $H_2$ and $H_3$ assumes then a
particularly simple form:
$${ \sum_{j\in\Lambda} [h_{j,j+1}+h_{j+1,j+2},[h_{j,j+1},
h_{j+1,j+2}]] =0}.\eqlabel\testcon $$
The derivation of this result is very simple.  Starting from
$$\eqalign{
H_3 =[B,H_2] =& \sumL (j-1)[h_{j-1,j},h_{j,j+1}] + (j+1)
[h_{j+1,j+2},h_{j,j+1}]\cr
=&~-\sumL ~[h_{j,j+1},h_{j+1,j+2}],\cr}\eq$$one then enforces
$$\eqalign{[H_3,H_2] =&\sum_j [[h_{j-2,j-1},h_{j-1,j}],h_{j,j+1}]+
[[h_{j-1,j},h_{j,j+1}],h_{j,j+1}]\cr +&~~
[[h_{j,j+1},h_{j+1,j+2}],h_{j,j+1}]+
[[h_{j+1,j+2},h_{j+2,j+3}],h_{j,j+1}]=0.\cr}\eq$$
Using
the Jacobi identity, the first term on the rhs
can be written in the form
$-[h_{j-1,j},h_{j,j+1}],h_{j,j+1}]$; with the shift $j\ra j+2$, it cancels
exactly the fourth term.
Then, by shifting $j$ by one unit in the second term, we
recover (\testcon).

{}From the Jacobi identity, this condition automatically ensures the existence
of
a second nontrivial conservation law $H_4$ which commutes with
$H_2$:
$$[H_4,H_2] = [[B,H_3],H_2] = -[[H_3,H_2],B] - [[H_2,B],H_3]=0.\eq$$ However,
showing that
$[H_4,H_3]=0$ and that higher charges commute with
$H_2$ requires additional information.{
For example, if it is known beforehand
that the commutant of $H_2$ is abelian,
(\testcon) actually implies the existence of an infinite tower of charges in
involution.}

\subsec{Relation between conjecture 2 and the Reshetikhin condition}

It is known (see [\ref{
Tetelman M G 1981  {\it Sov. Phys. JETP} {\bf 55} 306}\refname\Tet])
that the existence of such a
ladder operator is a direct consequence of the Yang-Baxter equation for
nearest-neighbor interacting chains
 for which the transfer matrix is a product of $R$
matrices (the so-called fundamental spin chains).  But the conjecture is
a priori independent of
the Yang-Baxter equation and in principle there could exists models satisfying
(\testcon) and not the Yang-Baxter equation.  Furthermore, (\testcon) is
easier to test that the Yang-Baxter equation
for the related R matrix.

Actually, for fundamental spin systems, (\testcon) can be viewed as a condition
for the matrix
$$R(\la) = P[I + \la H_2+ \calO(\la^2)]\eq$$
to be a solution of the Yang-Baxter equation.  This approach leads to the
condition
$$[h_{j,j+1}+h_{j+1,j+2},[h_{j,j+1},
h_{j+1,j+2}]] =X_{j,j+1} - X_{j+1,j+2}\eqlabel\Rcond$$
for some quantity $X$.  This relation first
appeared in [\ref{Kulish P P and
Sklyanin E K 1982 in {\it Integrable Quantum Field Theories} ed J Hietarinta
and C Montonen (Lecture Notes in Physics {\bf 151}, Springer Verlag) p.
61}\refname\KS]
(see eq. (3.20)) and it is attributed to Reshetikhin.  An explicit derivation
can be
found in [\ref{Kennedy T 1992 {\it J. Phys. A: Math. Gen}  {\bf 25}
2809}\refname\Ken].
The Reshetikhin condition (\Rcond) is
nothing but the local version of (\testcon).
It appears to be an anticipation of the boost construction of
conservation laws. It is also pointed out in
[\KS] that this equation is not satisfied by all integrable systems (which is
by now
understood from the fact that not all such systems are fundamental);
in particular this is the
case for the Hubbard model (in agreement with the conclusion in [\GMa]
 concerning the non-existence of a boost operator).

In the rest of this work we examine the existence of a third-order charge for a
number
of
models.
Some of the calculations
were performed using
``Mathematica".

\newsec{Example 1: spin-$1/2$ chains
with next-to-nearest neighbors interactions and bond
alternation}

\subsec{Definition of the model}

We consider a 10-parameter family of spin-1/2 models, which contain,
in addition to
XYZ-type interactions, next-to-nearest
neighbor interactions, bond alternation terms and a magnetic field
coupling term:
$$ H_2 =
\sum_{j\in \Lambda} \{ \sum_{a=x,y,z} [ \lambda_a \s^a_j \s^a_{j+1} +
\lambda\pr_a (-1)^j \s^a_j \s^a_{j+1}+
\lambda\bis_a \s^a_j \s^a_{j+2}] +h \s^z_j\}.\eqlabel\hfam$$
This is the most general spin-1/2 model with interaction range shorter
 then two lattice units admitting bond alternations.
To ensure translational invariance,
we assume that the  number of sites $N$ is even.
These models can be equivalently represented as nearest-neighbor
interactions on a  lattice $\Lambda'$,
whose bonds correspond to nonvanishing interactions
(see Fig. 1).

$$ {\epsfxsize= 10cm \epsfbox{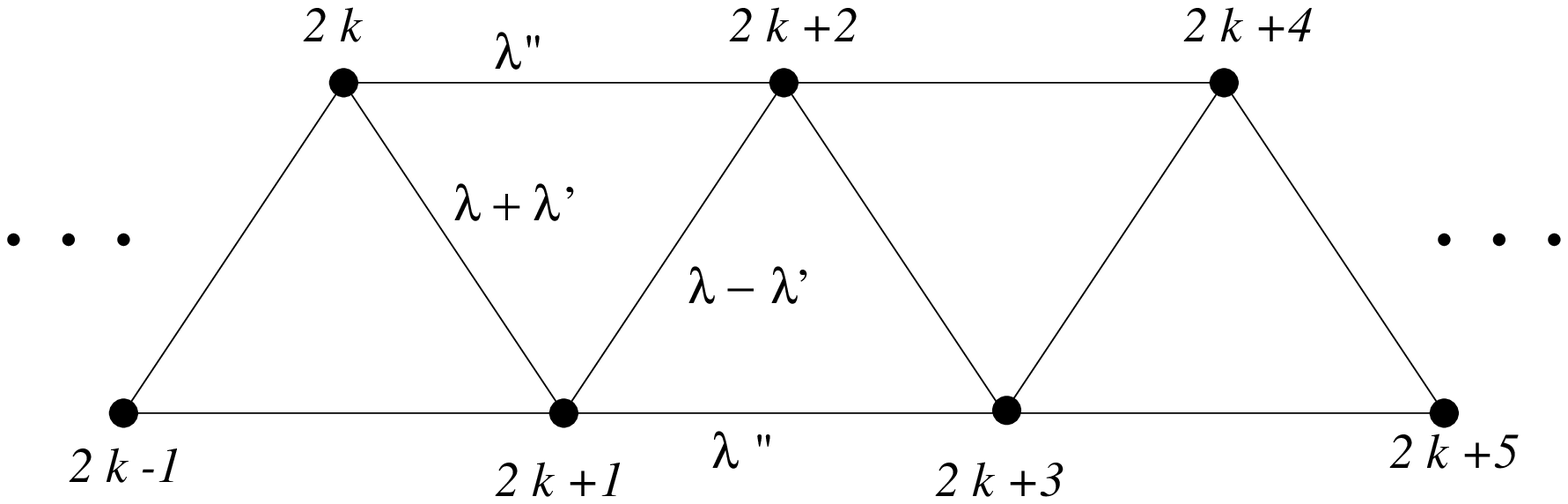}}\ \ \  $$

Fig. 1.
 The lattice $\Lambda'$ corresponding to the hamiltonian (\hfam), whose
bonds correspond to nonvanishing interactions.
\vskip 1cm

A lattice with such a ``railroad trestle" topology has been considered
in [\ref {Long M W and Siak S 1991
{\it J. Phys. C: Cond. Mat.} {\bf 3} 4901}\refname\RR] (in the
case where all the couplings are equal).
Notice  that the lattice in Fig. 1 corresponds also to a
generalization of (\hfam)
admitting  bond alternation for next-to-nearest neighbors.
A particular case of such bond alternation yields then
the ``sawtooth" topology, which has been shown to posses an
exact valence-bond ground state [\ref{ Long M W and Fehrenbacher R 1990
{\it J. Phys. C: Cond. Mat.} {\bf 2} 2787}].

The models in (\hfam) can be equivalently described by a hamiltonian of
the form
(\hamclass). The structure of (\hamclass) is recovered if we express
(\hfam) in terms of the variables:
$$S_i =\pmatrix{\s_{2i-1} \cr \s_{2i} }  {\quad \rm or}\quad
S_i =
\pmatrix
{\s_{2i}\cr \s_{2i+1} } .\eq$$

The family (\hfam) contains many interesting systems, including some that are
well known integrable models and some which are ``exactly solvable" in
some sense.
In particular, among the class of isotropic
(globally $su(2)$-invariant)  models satisfying
$ h=0$ and
$$\lambda_a=\lambda, \quad \lambda\pr_a=
\lambda\pr, \quad \lambda\bis_a=\lambda\bis,\eq$$
for all $a$,
 the following special cases are covered by  (\hfam):

\n  (i)
 the Heisenberg (XXX) model ($\lambda\pr=\lambda\bis=0$):
$$ H_2 =
\sumL \lambda \s^a_j \s^a_{j+1} ,\eq$$

\n (ii) the staggered XXX model ($\lambda\bis=0$):
$$ H_2 =
\sumL [\lambda+\lambda\pr (-1)^j] \s^a_j \s^a_{j+1} ,\eq$$

\n (iii) the alternating XXX model ($\lambda\bis=\la=0$):
$$ H_2 =
\sumL \lambda (-1)^j \s^a_j \s^a_{j+1} ,\eq$$

\n  (iv) the Majumdar-Ghosh model
[\ref{ Majumdar C K and Ghosh D K 1969 {\it J. Math. Phys.}
 {\bf 10} 1388, 1399}\refname\MG]:
($\lambda\pr=0,\lambda\bis=\hal~\lambda$):
$$ H_2 =
\sumL[
\lambda \s^a_j \s^a_{j+1} +
\hal \lambda\s^a_j \s^a_{j+2}]. \eq$$
As is well known this model possesses an exact valence-bond ground state.

For models invariant under global spin rotation around the
$z$-axis:
 $$\pmatrix{\s^x_i\cr \s^y_i} \to
 \pmatrix{\cos\alpha & \sin\alpha \cr -\sin\alpha&\cos\alpha}
 \pmatrix{\s^x_i\cr \s^y_i} ,\eqlabel\spinrot$$
 (in which case the global $su(2)$ invariance is broken down to $o(2)$), the
coupling constants satisfy:
$$\lambda_x=\lambda_y,\quad
\lambda\pr_x=\lambda\pr_y, \quad
\lambda\bis_x=\lambda\bis_y.\eq$$
Some interesting $o(2)$-symmetric models that can be
obtained from a specialization of (\hfam) are:

\n (v) the XXZ model ($\lambda\pr_a=\lambda\bis_a=0$):
$$ H_2 =
\sumL
[\lambda_x (\s^x_j \s^x_{j+1} + \s^y_j \s^y_{j+1}) +
\lambda_z \s^z_j \s^z_{j+1}],\eq$$

\n (vi)
the Lieb-Schultz-Matis model with alternating Heisenberg and Ising bonds
[\ref{Lieb E,  Schultz T  and  Mattis D 1961 {\it Ann. Phys}
{\bf 16}  407}\refname\LSM]:
$$\eqalign{
&\lambda\bis_a=0, \quad \lambda_x=\lambda_y=
\lambda\pr_x=\lambda\pr_y={1 \over 2}, \cr &
\lambda_z=(1+U)/2, \quad \lambda\pr_z=
(1-U)/2,} \eqlabel\LSMpar$$ with the hamiltonian
$$ H_2 =
\sumL
[ \s^a_{2j} \s^a_{2j+1} + U\s^z_{2j+1} \s^z_{2j+2} ]
,\eq$$
\n (vii)  the Hubbard model:
 $$ \eqalign{
&\lambda_x=\lambda_y=\lambda\pr_x=\lambda\pr_y=
\lambda\bis_z=0, \quad \lambda\bis_x=\lambda\bis_y=1,
\cr &  \lambda\pr_z=\pm ~ \lambda_z=U
,}\eq $$
with the hamiltonian
$${H_2=\sumL [
\s^+_{j}\s^-_{j+2} + \s^-_{j}\s^+_{j+2}+
    U [1+(-1)^j]\s^z_{j} \s^z_{j+1}].}\eqlabel\Hubeq$$
The  equivalence of (\Hubeq) (for a lattice with an even number of sites)
with the usual formulation
of the Hubbard model
$$H = \sumL [s^x_js^x_{j+1}+s^y_js^y_{j+1}+t^x_jt^x_{j+1}+
t^y_jt^y_{j+1}+Us^z_jt^z_{j}]\eqlabel\hubham$$
(where $s_i$ and  $t_i$ are two independent
sets of Pauli matrices at site $i$),
can be seen by
redefining the spin
variables in (\Hubeq) as $$s^a_j\to \s^a_{2j}, \quad
t^a_j\to \s^a_{2j+1}.\eqlabel\hubred$$

\subsec{Strategy of the test}

For the completely general anisotropic case,  a natural candidate for the
third order charge involves only
nearest neighbor interactions on the the lattice in Fig. 1 and has the form:
\eqn\Hthfaman{\eqalign{H_3 =&\sumL \sum_{a_i=x,y,z} [( \alpha^{(1)}_{a_1a_2a_3}
+ (-1)^j \alpha\pr^{(1)}_{a_1a_2a_3})
\s^{a_1}_j \s^{a_2}_{j+1} \s^{a_3}_{j+2} \cr &~~+
(\alpha^{(2)}_{a_1a_2a_3} +(-1)^j \alpha\pr^{(2)}_{a_1a_2a_3})
\s^{a_1}_j\s^{a_2}_{j+2}\s^{a_3}_{j+3}\cr &~~+
 ({\alpha^{(3)}}_{a_1a_2a_3} +(-1)^j \alpha\pr^{(3)}_
{a_1a_2a_3}) \s^{a_1}_{j}\s^{a_2}_{j+1}\s^{a_3}_{j+3} \cr &~~+
(\alpha^{(4)}_{a_1a_2a_3} +(-1)^j \alpha\pr^{(4)}_{a_1a_2a_3})
\s^{a_1}_{j}\s^{a_2}_{j+2}\s^{a_3}_{j+4} ]}
  }
where $\alpha^{(i)}_{a_1a_2a_3}$
and $\alpha'^{(i)}_{a_1a_2a_3}$
are arbitrary coefficients.
We search for integrable systems by imposing the
condition of commutativity of $H_3$ with the hamiltonian. This leads
to an over-determined system of equations for the set of parameters in
$H_3$. This system has nontrivial solutions only for special
values of the parameters of the hamiltonian.
However, the analysis of this system is very cumbersome, since
in absence of additional symmetries, \Hthfaman\
contains 216 free parameters!
Henceforth, we consider only three special situations, in which the analysis
simplifies significantly: the isotropic case, the $o(2)$-symmetric case,
 and  the anisotropic case without bond alternation nor next-to-nearest
neighbor
interaction - that is, the XYZh model.


One might consider other candidate charges, involving other triples of sites
than those in \Hthfaman. However,
suppose that one cluster different from those appearing in
\Hthfaman\ is introduced. Then (considering an infinite chain),
to cancel the new terms arising in the commutator with the hamiltonian,
an infinite sequence of other clusters would have to be added,
with the distance
between spins in such clusters growing arbitrarily.
This would violate the requirement of locality.
(The above  reasoning is not so obvious however
when some of the couplings vanish.)

\subsec{The XYZh model}

We first consider
(\hfam) in absence of bond alternation and  next-to-nearest neighbor terms.
The most
general candidate charge $H_3$,
coupling three nearest-neighbor spins, is:
\eqn\Hthxyah{H_3 =\sumL \sum_{a_i=x,y,z}  \alpha_{a_1a_2a_3}
\s^{a_1}_j \s^{a_2}_{j+1} \s^{a_3}_{j+2},}
which contains 27 free parameters.
By enforcing the commutativity of this candidate $H_3$ with the hamiltonian,
we found that if $h\ne 0$ and no two couplings are equal,
the only solution is: $\alpha_{a_1a_2a_3}=0$ for all triples
$a_1a_2a_3$.
Thus there is no  nontrivial charge $H_3$ for the anisotropic XYZ chain in
a nonzero magnetic field. This  suggests thus that the XYZh model
is nonintegrable, in agreement with the fact
that the Bethe Ansatz solution  for the XYZ model
cannot be generalized to the case with
a nontrivial magnetic field.

\subsec{The isotropic case}
In the isotropic case, the most general third order charge involving only
the nearest neighbors on the lattice in Fig. 1 has the following form:
\eqn\Hthfam{\eqalign{H_3 =&\sum_{j\in\Lambda} \epsilon_{abc}
[( \alpha + (-1)^j \alpha\pr)
\s^a_j \s^b_{j+1} \s^c_{j+2} \cr &+
(\beta +(-1)^j \beta\pr) \s^a_{j}\s^b_{j+2}\s^c_{j+3} +
 (\tilde \beta +(-1)^j \tilde \beta\pr) \s^a_{j}\s^b_{j+1}\s^c_{j+3} +\cr &
(\gamma +(-1)^j \gamma\pr) \s^a_{j}\s^b_{j+2}\s^c_{j+4} ],}
  }
where $\alpha,\beta,\tilde{\beta},\gamma$ and their primed
variants are arbitrary parameters.
Solving the commutativity condition $[H_2, H_3]=0$,
we found that (for  $N\ge 8$)
nontrivial solutions (for which not all parameters of $H_3$ are
equal to zero) exist only in two cases:
for the XXX model ($\lambda\pr=\lambda\bis$),
with the solution $\alpha\ne 0$, $\beta=\beta'=\tilde{\beta}=
\tilde{\beta}'=\gamma=\gamma'=0$,
i.e. $$ H_3=\sumL \alpha \epsilon_{abc}
\s^a_j \s^b_{j+1} \s^c_{j+2} ,\eqlabel\XXXsol$$
and for $\lambda=\lambda\pr$, which
corresponds to two decoupled XXX models
on the even and odd sublattices
(the solution is then $\gamma,\gamma\pr\ne 0$ and all other
parameters of $H_3$ equal to zero).

These  results  suggest
that the only integrable isotropic models
within the family (\hfam) are of the XXX-type. In particular,
the Majumdar-Ghosh model,
the alternating XXX model, and the staggered XXX   model\foot{Note
that in the continuous limit the alternating term corresponds
to $\tr~ g$ in the WZNW model
[\ref{ Affleck I 1985 {\it Phys. Rev. Lett.} {\bf 55} 1355;
1986 {\it Nucl. Phys.}  {\bf B265} 409 }\refname\Affleck] with a level one
affine su(2) spectrum generating algebra.
$g$ stands for the basic field in the WZNW model, a $2\times 2$ matrix in the
$su(2)$ case.
$\tr ~g$ turns out to be an integrable perturbation of this WZNW model
[\ref{Kobayashi K
and Uematsu T 1990 {\it Mod. Phys. Lett.} {\bf A5} 2515; Leblanc M,  Mann R B
and
Zheng H B {\it Nucl. Phys.} {\bf B349} 220}]. However, the staggered XXX model
in the
continuous limit gives the WZNW model perturbed by both $\tr ~g$ and the
marginal
current-current term.  Together, these two perturbations are incompatible with
integrability. Note also that for $\la=\pm\la'$ the staggered XXX model
degenerates into a pathological ``chopped XXX" chain consisting of
disjoint bonds (see section 3.5 for a discussion of a similar
case).}
 all fail the test of the existence of $H_3$ and thus seem to be nonintegrable.

For $N<8$ there exist nontrivial solutions for $H_3$ of the form
\Hthfam. Let us illustrate this in the case
$\a'=0$, when the hamiltonian (\hfam)
contains only  nearest and next-to-nearest neighbors,
$$H_2=H+\mu H',\eqlabel\gennn$$
where $\mu$ is an arbitrary parameter, and $H$ and $H'$ are given by (\Hxxx)
 and (\Hpxxx) respectively.
For $N=5$ we find that the XXX charge (\XXXsol) commutes with (\gennn).
This is a consequence of the ``accidental" identity
(\fiveid) holding for $N=5$. In other words, for $N=5$
the next-to-nearest interaction is a ``nonlocal" charge,
(related to the square of the total spin), which commutes with all the
XXX charges. For $N>5$ this solution disappears (the three-spin XXX
charge no longer commutes with the hamiltonian (\gennn)). Similarly, for
$N=6$ there is a one parameter family of solutions
$$ H_3=H_3^{XXX} + \nu F_{3,1} -1/3 ~(\mu+\mu\nu-\nu) F^s_{3,2},\eq$$
where $\nu$ is an arbitrary parameter, and\foot{
$F^s_{3,2}$ is the symmetric part of the quantity
$F_{3,2}$ introduced in [\GMa].}
$$ F_{3,1}=\sumL \epsilon^{abc}[ \s_j^a \s_{j+2}^b \s^c_{j+3} +
\s_j^a \s_{j+1}^b \s^c_{j+3}], \eq$$
$$ F^s_{3,2}=\sumL \epsilon^{abc} \s_j^a \s_{j+2}^b \s^c_{j+4}. \eq$$
For $N=7$ there is another solution:
$$H_3=H_3^{XXX}+\mu F_{3,1} -\mu^2 F^s_{3,2}.\eq $$
Similar accidental nonlocal three-spin charges,
whose form changes with  $N$
exist in fact for all $N$, but it is only for
$N<8$ that they ``look local", i.e. can be put in the form \Hthfam.

\subsec {The $o(2)$-invariant case}

We  are searching again for a nonvanishing charge $H_3$ of the form \Hthfaman.
The requirement of the $o(2)$ invariance imposes a number of restrictions
on the parameters
$\alpha^{(i)}_{a_1a_2a_3}$ and
$\alpha'^{(i)}_{a_1a_2a_3}$:
$$ \eqalign{&
\al_{yxy}= \al_{zxz}= \al_{xyx}= \al_{zyz}= \al_{yyx}= \al_{xyy}=
\al_{zzx}= \cr
&\al_{xzz}= \al_{zzy}= \al_{yzz}= \al_{xxy}= \al_{yxx}= \al_{xxx}=
 \al_{yyy}=0,}\eq$$
and
$$\eqalign{&
\al_{yxz}= -\al_{xyz},
\al_{zyx}= -\al_{zxy},
\al_{xzy}= -\al_{yzx},\cr &
\al_{yzy}= \al_{xzx},
\al_{zxx}= \al_{zyy},
\al_{xxz}= \al_{yyz}.}\eq$$
As a result, the number of free parameters in $H_3$
is now decreased to 56.
We then check whether there are nontrivial $H_3$ of this form
commuting with the hamiltonian. For $N=8$ for example, this leads
to a linear system of roughly 240 equations with 56 unknowns,
which depends on the 6 free parameters of the hamiltonian
($\a,\c, \a',\c',\a'', \c''$).
Again one is looking for the values of these parameters allowing
for the existence of a nontrivial solution (persisting when $N$ is increased).
This system is best analyzed with a computer. The results are given
below.\foot{Note that trivial $o(2)$-symmetric models with
hamiltonians involving only the $z$-components of the spin variables at each
site
are explicitly excluded from consideration.}

If $\a''$ is not zero, a nontrivial solution for $H_3$ exists only
in two cases:

\n (a)
for two decoupled XXZ models on two disjoint (even and odd) sublattices:
($\a=\a'=\c=\c'$),


\n (b) the Hubbard model: ($\a=\a'=\c''=0, \c=\pm \c'$).
The solution for the charge $H_3$ is then
$$\eqalign{
H_3=& \sumL [
2 \c'( \s^y_j\s^z_{j+1}\s^x_{j+2} -\s^x_j\s^z_{j+1}\s^y_{j+2})+
\a''(\s^y_j\s^z_{j+2}\s^x_{j+4} -\s^x_j\s^z_{j+2}\s^y_{j+4})
\cr &+
\c' (1-(-)^j)
 ( \s^x_j\s^y_{j+2}\s^z_{j+3}
-\s^y_j\s^x_{j+2}\s^z_{j+3}  \cr &
- \s^z_j\s^y_{j+1}\s^x_{j+3}
+ \s^z_j\s^x_{j+1}\s^y_{j+3})
]
 ,}\eq$$
which can be translated, using (\hubred),
into the usual expression for the
third-order charge in the Hubbard model [\Shas, \GMa].

If $\a''$ is zero there are more possibilities. Nontrivial solutions
for $H_3$ exist in four cases. These are:

\n (c) the XXZ model: $\a'=\c'=\c''=0$,

\n (d) the staggered XX model: $\c=\c'=\c''=0$,

\n (e) the staggered XXZ model: $\a=\c'=\c''=0$,

\n (f) the model with alternating XXZ and Ising bonds: $\a=\pm \a'$.


The three-spin charge obtained for the XXZ model is:
$$\eqalign{H_3=&\sumL[
 \c (
\s^x_j\s^y_{j+1}\s^z_{j+2} -
\s^y_j\s^x_{j+1}\s^z_{j+2}+
\s^z_j\s^x_{j+1}\s^y_{j+2} -
\s^z_j\s^y_{j+1}\s^x_{j+2} )\cr &+
\la_x(\s_j^x \s_{j+1}^z\s_{j+2}^y-\s_j^y\s_{j+1}^z\s_{j+2}^x)],}
\eq$$
in agreement with [\Lush,\GMa].

For the staggered XX model
\eqn\stXX{ H_2=\sumL (\la + \la\pr (-1)^j)(\s^x_j\s_{j+1}^x+\s_j^y\s_{j+1}^y),}
the three-spin charge obtained via our test
is identical to the XX charge, that is
$$H_3=\sumL \alpha (\s_j^x \s_{j+1}^z\s_{j+2}^x+\s_j^y\s_{j+1}^z\s_{j+2}^y)
+\beta (\s_j^x
\s_{j+1}^z\s_{j+2}^y-\s_j^y\s_{j+1}^z\s_{j+2}^x),\eqlabel\hteststXX$$
where $\alpha$ and $\beta$ are arbitrary coefficients.
Another indication of integrability of this model is provided by
its continuum limit, which corresponds to a theory of free massive fermions
[\ref{Affleck I 1989 in {\it Fields, Strings and Critical
Phenomena} ed E Br\'ezin and J Zinn-Justin
(Elsevier)}\refname\AffleckLect].
Since we are not aware of an
explicit proof of the integrability of the lattice model \stXX\
in the literature,
we present a simple direct proof
by exhibiting a family of mutually commuting conservation laws.
These can all be expressed in terms of the densities:
$$e_{n,j}^{\alpha\beta}=
\s^\alpha_j\s^z_{j+1}\dots \s^z_{j+n-2}\s^\beta_{j+n-1},\eq$$
defined for $n\ge2$.
In terms of these quantities, the scalar and pseudoscalar
conserved charges [\GMa] of the XX model are:
\eqn\hxypl{ \eqalign{ h^{(+)}_n &=
 \sumL e_{n,j}^{xx} +e_{n,j}^{yy}\quad\quad n~{\rm even}, \cr
&=\sumL
 e_{n,j}^{xy} -e_{n,j}^{yx}\quad\quad n~{\rm odd},
}}and
\eqn\hxym{ \eqalign{ h^{(-)}_n&=
\sumL e_{n,j}^{xy}-e_{n,j}^{yx}\quad\quad n~{\rm even}, \cr &=
\sumL e_{n,j}^{xx}+e_{n,j}^{yy}\quad\quad n~{\rm odd}.  }}
We also define
$$ \eqalign{
k_n^{(+)}=&\sumL (-1)^j (e_{n,j}^{xx}+e_{n,j}^{yy}),\cr
k_n^{(-)}=&\sumL (-1)^j (e_{n,j}^{xy}-e_{n,j}^{yx}).}\eq$$
The conserved charges of the staggered XX model contain
two families $H_n^{(\pm)}$. For $n$ odd these charges
coincide with the XX charges:
$$ \eqalign{
H_{2m+1}^{(+)}=& h^{(+)}_{2m+1}
,\cr
H_{2m+1}^{(-)}=& h^{(-)}_{2m+1}
\quad .\cr}\eqlabel\allch
$$
For $n$ even, the conserved charges for \stXX\ are:
$$ \eqalign{ H^{(+)}_{2m}=& \la h_{2m}^{(+)}+\la' k_{2m}^{(+)}+
\la h_{2m-2}^{(+)}-\la' k_{2m-2}^{(+)}, \cr
H^{(-)}_{2m}=& \la h_{2m}^{(-)}+\la' k_{2m}^{(-)}+
\la h_{2m-2}^{(-)}-\la' k_{2m-2}^{(-)}.}\eq$$
Mutual commutativity of the charges $H_n^{(\pm)}$ as well as their commutation
with the staggered XX hamiltonian (for $|\Lambda|$ even) can be verified
directly as in [\GMa]. Note also that the boost operator,
$$ B=
{1\over {2i}} \sumL j[ \la (\s^x_j\s^x_{j+1}+\s^y_{j}\s^y_{j+1})+
 \la' (-1)^j (\s^x_j\s^x_{j+1}+\s^y_{j}\s^y_{j+1})], \eq$$
has the ladder property: acting on \stXX\ it produces
the scalar part of (\hteststXX).

Under a spin rotation by $\pi/2$
around the $z$-axis restricted to odd sites, i.e.:
$$\eqalign{
&\s_{2j+1}^x\to\s_{2j+1}^y,~\quad \s^y_{2j+1}\to -\s^x_{2j+1},\cr
&\s_{2j}^x\to\s_{2j}^x,\quad\quad\quad\quad
 \s^y_{2j}\to\s^y_{2j},\cr}\eqlabel\DMdual$$
the alternating part of the staggered XX hamiltonian
transforms into the two-spin pseudoscalar charge $h_2^{(-)}$ of the XX model,
$$  \sum_{j\in\Lambda} (-1)^j
(\s^x_j \s^x_{j+1}+ \s^y_j \s^y_{j+1}) \to h_2^{(-)},\eq$$
where $$h_2^{(-)}=\sum_{j\in\Lambda}[ \s^x_j \s^y_{j+1}- \s^y_j \s^x_{j+1}].
\eq $$
This is a special case of the Dzyaloshinski-Moriya interaction
[\ref{ Dzyaloshinski I E 1958 {\it J. Phys. Chem. Solids} {\bf 4} 241;
Moriya T 1969 {\it Phys. Rev. Lett} {\bf 4} 228}].
Notice that the transformation (\DMdual) can be interpreted as a duality in the
sense of (\sdham)  if we define
$\tilde\s_{2j+1}^x= i \s_{2j+1}^y,~\tilde\s^y_{2j+1}=
-i \s^x_{2j+1}$  (where the factor $i$ has been introduced in
order to have $\tilde{\tilde \s}=\s$). Then $\tilde h_2^{(+)}=i k_2^{(-)}$ and
$\tilde h_2^{(-)}=i k_2^{(+)}$. One may then consider a general
hamiltonian:
$$  H=\la_1 h_2^{(+)}+ \la_2  h_2^{(-)} +
\la_3 k_2^{(-)}+ \la_4  k_2^{(+)},\eqlabel\gensd$$
where $\la_1, \la_2,\la_3$ and $\la_4$ are arbitrary constants.
This hamiltonian is integrable,
as can be seen from the existence of an infinite family of conservation laws,
given again (for $n$ odd) by (\allch).
Notice that (\gensd) is self-dual
for $\la_1 \la_4=\la_2 \la_3$; however,
it does not satisfy the
Dolan-Grady sufficient integrability condition (\sdeq).

For the staggered XXZ model, defined by the hamiltonian:
\eqn\stXXZ{H_2=\sumL [\la\pr_x (-1)^j (\s_j^x\s_{j+1}^x+\s_j^y\s_{j+1}^y)+
\c \s_j^z\s_{j+1}^z],}
the three-spin charge obtained from the test has the following form:
$$\eqalign{H_3=&\sumL
[ \c (-1)^j (
\s^x_j\s^y_{j+1}\s^z_{j+2} -
\s^y_j\s^x_{j+1}\s^z_{j+2}+
\s^z_j\s^x_{j+1}\s^y_{j+2} -
\s^z_j\s^y_{j+1}\s^x_{j+2} )\cr &+
\la_x\pr (\s_j^x \s_{j+1}^z\s_{j+2}^y-\s_j^y\s_{j+1}^z\s_{j+2}^x)].}
\eq$$
This charge can be also obtained using the boost operator
as $[B, H_2]$.
Note that the transformation (\DMdual) establishes the
equivalence of the staggered XXZ chain with the model
$$H_2=\sum_{j\in\Lambda} [\s^x_j \s^y_{j+1}- \s^y_j \s^x_{j+1}+
\c \s^z_j \s^z_{j+1}], \eq $$
which is a particular case of the XXZ model with
Dzyaloshinski-Moriya interaction:
$$H_2=\sum_{j\in\Lambda}
J_x(\s^x_j \s^x_{j+1}+ \s^y_j \s^y_{j+1})+
D(\s^x_j \s^y_{j+1}- \s^y_j \s^x_{j+1})+
J_z \s^z_j \s^z_{j+1} ,\eqlabel\DMXXZ $$
where $J_x, J_z$ and $D$ are arbitrary parameters.
Integrability of \stXXZ\
 follows then from the integrability of (\DMXXZ),
which has been proven in
[\ref{Alcaraz F C and Wrzesinski W F 1990 {\it J. Stat. Phys.}
{\bf 58} 45}].
Note also that the model (\DMXXZ)
is not an $o(2)$-invariant one,
and by a  spin rotation (\spinrot) with a suitably chosen
angle $\alpha$ it may be transformed
into the anisotropic Dzyaloshinski-Moriya system:
$$H_2=\sum_{j\in\Lambda}
[D_x\s^x_j \s^y_{j+1} +D_y \s^y_j \s^x_{j+1}+
D_z \s^z_j \s^z_{j+1}] .\eqlabel\DMan$$

The model with alternating XXZ and Ising bonds,
$$ \eqalign{
H_2 =&
\sumL
[(\a+\a') (\s^x_{2j} \s^x_{2j+1}+  \s^y_{2j} \s^y_{2j+1} )
+ (\c+\c') \s^z_{2j} \s^z_{2j+1} \cr & +
 (\c-\c') \s^z_{2j+1} \s^z_{2j+2} +\c'' \s^z_{j} \s^z_{j+2}]
 ,}\eqlabel\genLSM$$
which is a slight generalization of the Lieb-Schultz-Mattis model,
presents certain peculiarities. The model has been diagonalized
in [\LSM] (for
$\lambda\bis_z=0,
\lambda_x=\lambda\pr_x={1 \over 2}, \lambda_z=(1+U)/2, \lambda\pr_z=
(1-U)/2 $).
 As observed in [\LSM], a convenient basis is
provided by the eigenstates of ${\bf L}_j=\S_{2j}+\S_{2j+1}$.
Consider then subspaces of the space of states corresponding to a
particular sequence $\{M_j\}$ of the eigenvalues
of the third component of the ${\bf L}_j$'s.
Since the Lieb-Schultz-Mattis type hamiltonian (\genLSM)
commutes with each of the ${\bf L}_j$'s it
does not mix different subspaces; in other words it is block-diagonal in
this basis, and it can be diagonalized separately in each subspace.
The projection operators onto the subspaces corresponding to different
sequences $\{M_j\}$ provide then a
mutually commuting set of operators, all commuting with
the hamiltonian.
Therefore, (\genLSM) satisfies the
definition of integrability given in the introduction; admittedly
the nature of these conserved charges appears  a bit unusual
and somewhat trivial.

The explicit form $H_3$ charge found via the the test is:
$$ H_3=\sumL [\alpha L^z_j \s^z_{2j+2} \s^z_{2j+3}+
\beta   \s^z_{2j} \s^z_{2j+1} L^z_{j+1}],\eqlabel\hzz$$
where $\alpha$ and $\beta$ are arbitrary coefficients.
Interestingly, not only
does the above sum  commute with (\genLSM),
but each term in it is separately conserved.
Clearly, all these terms can be expressed as
linear combinations of the
projection operators discussed above.
Let $P_j^{(m_j)}$ denote a projection onto the states with $M_j=m_j$.
Then $\s^z_{2j} \s^z_{2j+1}={1 \over 4} (P_j^{(1)}+P_j^{(-1)}-P^{(0)}_j) $ and
$L_j^z=P_j^{(1)} - P_j^{(-1)}$. In particular, in the
$2^{N/2}$-dimensional subspace where all the $M_i$'s
are zero (which is the sector containing the vacuum),
(\hzz) vanishes.

Clearly, the block-diagonal nature of the hamiltonian
(and hence the existence of the set of commuting projections) is
preserved by the addition to (\genLSM)
of an arbitrary interaction involving only the
$z$-components of the spin variables.
Finally, we note that a  particular case
of (\genLSM) with $\c-\c'= \c''=0$ provides
a pathological ``chopped XXZ" system, consisting of $N/2$ disjoint
XXZ bonds, with all neighboring links trivially commuting.

Summing up, it appears that the apart from the models (a)-(f)
described above, all other $o(2)$-symmetric models within the family (\hfam)
are nonintegrable.
Among the integrable models, there are  three situations
in which the first moment of the hamiltonian (the boost) acts as a
ladder operator for conservation laws:
the XXZ chain (cases (a) and (c)), the staggered XX model (d),
and the staggered XXZ chain (e).


\newsec{Example 2: isotropic spin-$1$ chains}
Consider now a class of isotropic spin-1 chains with nearest neighbor
interactions. The most general hamiltonian contains a bilinear and a
biquadratic term:
\eqn\hspinone{
H_2(\beta)=\sum_{j\in\Lambda}[ {S^a_j S^a_{j+1} }+
\beta ( {S^a_j S^a_{j+1}})^2],}
where $S^a_j$'s  are the $su(2)$ spin-1 matrices,
 acting nontrivially only on the
$j$-th factor of the Hilbert space $\bigotimes_j \CC^3$.
For convenience  we choose the representation in which $S^z$ is diagonal,
i.e. $$ S^x={1\over {\sqrt{2} }} \pmatrix{0&1&0\cr 1&0&1\cr 0&1&0},
	S^y={i\over {\sqrt{2} }} \pmatrix{0 &-1&0\cr 1&0 &-1\cr 0&1&0},
	S^z=\pmatrix{ 1&0&0\cr 0&0&0\cr 0&0 &-1}.\eq$$
Using the identity:
$$ (S^a_j S^a_k)^2={1\over 4} D^{ab}_{j} D^{ab}_k -{1\over 2}S^a_j S^a_k,
\eq$$
where
$$ D^{ab}_j\equiv S^a_j S^b_j+ S^b_j S^a_j \eq$$
the hamiltonian can expressed as:
\eqn\hspinonem{ H_2(\beta)=\sum_{j \in \Lambda} [(1-{\beta\over 2}) {S^a_j
S^a_{j+1} }+
{\beta \over 4} D^{ab}_j D^{ab}_{j+1}].}
The boost operator yields the following candidate for the $H_3$ charge:
\eqn\htspone{ \eqalign {H_3(\beta)=& \sum_{j\in \Lambda} \epsilon^{abc}[
(-2+2\beta-{\beta^2 \over 2} ) S^a_j S^b_{j+1} S^c_{j+2}
-{\beta^2 \over 2} D^{ad}_j S^b_{j+1} D^{cd}_{j+2}
\cr &+
(-\beta+{\beta^2\over 2}) (D^{ad}_j D^{db}_{j+1} S^c_{j+2}+
S^a_j D^{bd}_{j+1}D^{dc}_{j+2})
].}}
The commutator $[ H_2(\beta), H_3(\beta)]$
vanishes only for $\beta=\pm 1$ or $\beta=\infty$;
these  cases
have been already identified as integrable in the literature.
For  $\beta=-1$, \hspinone\  reduces to  the isotropic version of the
Fateev-Zamolodchikov chain,
associated with the 19-vertex model [\ref{
Zamolodchikov A B and Fateev V A 1980 {\it J. Sov. Phys.}
 {\bf 32} 298}], whose
integrability follows from directly from the Yang-Baxter equation.
For  $\beta=1$, \hspinone\ describes the Sutherland $su(3)$ symmetric chain
[\ref{Sutherland B 1975  {\it Phys. Rev.}
{\bf 12} 3795}\refname\Suth],
 whose hamiltonian can be rewritten in terms of the
Gell-Mann matrices $t^a$:
\eqn\htwosponegm{H_2(1) \sim \sum_{j\in \Lambda} t^a_{j}t^a_{j+1}.}
It can be solved by the nested Bethe Ansatz [\Suth] and it is also
related directly to the Yang-Baxter equation. Note that in this case, $H_3(1)$
can also be written in
the form [\GMa]
 \eqn\hsunthr{H_3=\sumL f^{abc} t^a_j t^b_{j+1} t^c_{j+2}.}
In the limit $\beta\to\infty$, \hspinone\ reduces to a system with purely
biquadratic
interactions,
whose integrability
has previously been established in [\Suth, \ref{Parkinson J B
1987 {\it J. Phys. C: Solid State Phys.} {\bf 20} L1029; 1988
{\it J. Phys. C: Solid State Phys.} {\bf 21} 3793;
Barber M N, Batchelor M T 1989
{\it Phys. Rev.} {\bf B40} 4621}\refname\BB].

For the spin-1 models \hspinone\  with finite $\beta\ne \pm 1 $,
the boost operator does not produce a conserved quantity.
Furthermore, as we show below, there exists no nontrivial
local conserved charge involving up to three nearest-neighbors.
The most general expression for such a charge would have the form:
\eqn\hgene{ \eqalign {H_3=& \sum_{j\in \Lambda} [ a_1
\epsilon^{abc}
S^a_j S^b_{j+1} S^c_{j+2} +
a_2 \epsilon^{abc} D^{ad}_j D^{db}_{j+1} S^c_{j+2}+
a_3 \epsilon^{abc} S^a_j D^{bd}_{j+1}D^{dc}_{j+2}\cr &~~+
a_4 \epsilon^{abc}
D^{ad}_j S^b_{j+1} D^{cd}_{j+2} +
a_5D^{ab}_j D^{bc}_{j+1} D^{ac}_{j+2}+
a_6  S^a_j S^b_{j+1} D^{ab}_{j+2} \cr&~~+
a_7  D^{ab}_j S^a_{j+1} S^b_{j+2} +
a_8 S^a_j D^{ab}_{j+1} S^b_{j+2} +
a_9 S^a_j S^a_{j+1} +
a_{10} ( {S^a_j S^a_{j+1}})^2]
.}}
where the $a_i$'s are undetermined coefficients.
Enforcing the commutativity of $H_3$ with the hamiltonian,
we obtain as usual a number of
constraints on these coefficients. In particular, the
vanishing of the four-spin terms in this commutator requires:
$$ a_2 - a_3= a_5=a_6=a_7=a_8=0,\eq$$
and
$$\eqalign{
&a_2=a_4=0\text{if}\beta=0,\cr
&a_1=(1+4/{\beta^2}-4/{\beta})a_4,\quad
a_2=(-1+2/\beta)a_4\text{if}\beta\not=0.\cr}\eq
$$
The above conditions means that in order for $H_3$ to commute with $H_2$, its
three-spin part must be proportional to the commutator of the boost and
$H_2$ (both for $\beta=0$ and $\beta\ne 0$).
The vanishing of the terms with two and three spins in the
commutator imposes two further restrictions.
First, the two-spin part of $H_3$ must be proportional to
the hamiltonian $H_2$. Second, unless $\beta^2=1$ or $\beta=\infty$ the
three-spin part of $H_3$ must be trivial.

The nonexistence of $H_3$ for \hspinone\ with finite $\beta\ne\pm 1$,
suggests that all these models are nonintegrable.
In particular, the bilinear system ($\beta=0$),
as well as the one with $\beta=1/3$, for which there exists an exact
valence-bond ground state [\ref{Affleck I, Kennedy T,  Lieb E H and  Tasaki H
1988 {\it Comm. Math. Phys.} {\bf 115} 477}\refname\AKLT],
all fail the above 
test.

It should be added that the necessary condition (\Rcond)
for having a quantum chain related to a solution of the
Yang-Baxter equation has been examined
for isotropic spin-$s$ chains with $s<14$ in [\Ken,
\ref{ Batchelor M T and Yung C M 1994
{\it Integrable $su(2)$-invariant spin chains}
preprint MRR-032-94 (cond-mat 9406072)}\refname\BaYu].
In particular, these authors found that the only spin-1 systems satisfying
(\Rcond) are $\beta=\pm 1$ and $\beta=\infty$. As we have discussed above,
this
implies that for other values of $\beta$  the hamiltonian
\hspinone\ cannot be a fundamental model. But a priori there
might exists integrable but non-fundamental models for $\beta\ne \pm 1,\infty$.
Our results,
showing that there is no nontrivial three-spin charge provide a much stronger
evidence for the
non-integrability of all the isotropic spin-1 models
with $\beta\ne \pm 1,\infty$. It would be interesting to
perform a similar analysis for $s>1$;  such analysis is
however much more complicated for higher $s$.

We end this section with a short remark on the general spin-$s$ bilinear
system.
Is it possible that this system is, by a bizarre accident,
an integrable fundamental model for some values of $s$?
This has been answered negatively in [\BaYu]
for all $s<100$,   using computer algebra.
Here we present a simple
calculation showing that for the bilinear systems,
the boost operator can never produce a
conserved quantity for
$s\ne 1/2$,
which thus excludes the possibility
of such an accident.
Acting on the hamiltonian, the boost operator
generates a candidate for the three-spin charge of the form:
$$ H_3= \sumL \epsilon^{abc} S^a_j S^b_{j+1} S^c_{j+2}. \eq$$
The commutator of this quantity with the hamiltonian is then
$$[H_3, H_2]=\sumL
[D^{ab}_j S^a_{j+1} S^b_{j+2}
-D^{aa}_j S^b_{j+1} S^b_{j+2}
+S^{a}_j S^a_{j+1} D^{bb}_{j+2}
-S^{a}_j S^b_{j+1} D^{ab}_{j+2}].\eq$$
The vanishing of this sum requires that
all the terms containing spins in an arbitrary cluster
vanish separately, which is not the case in general. The sum
can vanish only if the terms cancel two by two, which is possible
only if
$D^{ab}\sim \delta^{ab}$, a condition which is not true unless $s=1/2$.
Therefore, the bilinear Heisenberg chain is an integrable fundamental model
only for $s=1/2$.


\newsec{ Concluding remarks}

The simple integrability test considered in this work appears to
be applicable to a rather general class of quantum chains.
However, the range of applicability of the conjectures 1 and 2 has not
yet been determined rigorously.
Regardless of that, this simple method seems to be
a useful heuristic tool. It may be worthwhile to use it to try to
identify integrable models within other physically relevant
families of models.
In particular, the
completely anisotropic case of the 10-parameter family considered in
section 3 remains to be studied in detail. However, we do not
expect that
it will reveal new types of integrable
systems,  beyond anisotropic generalizations of the systems found in  the
$o(2)$-symmetric case. In particular, we expect that
the XYZ chain, the staggered XY model, the staggered XYZ chain (with the
$xy$ part alternating in sign), equivalent to the
Dzyaloshinski-Moriya system (\DMan),
a generalized Lieb-Schultz-Mattis model, and a generalized Hubbard model
(consisting of two copies of the $XY$ chain interacting along their $z$
components) are, up to a relabeling of
variables, the only nontrivial
integrable anisotropic models within the family (\hfam).

Conjecture 1  could be generalized in a natural
way to include even models with long-range interactions
[\ref{Haldane F D 1988 {\it Phys. Rev. Lett.}
{\bf 30} 635; Shastry B S 1988 {\it Phys. Rev. Lett.} {\bf
30} 639;
Inozemtsev V I 1990 {\it J. Stat. Phys} {\bf 59} 1143
}], for which there seems to be
no ladder operator.  We would
 then require that if a hamiltonian $H_2$ is
given by a sum of two-spin  interactions, there should
exist a conserved three-spin charge $H_3$.
Observe that for models with long range interactions, the
leading term in $H_n$ is also characteristic: although the $n$
interacting spins in the
leading term are no longer adjacent, the prefactor specifying
the interaction of these $n$
spins is distinctive.    But for these models there can exist
independent non-local charges
 also for finite chains (see e.g. [\ref{Haldane F D,
Ha Z N C,  Talstra J C,  Bernard D and
Pasquier D 1992 {\it Phys. Rev. Lett.} {\bf 69} 2021}\refname\HHTBP])
 and again it could be less obvious at first sight to assert that
a three-spin quantity is not a product of lower order charges.
 The  analysis of such models will be reported elsewhere.

Finally, we mention a completely different integrability test for spin
chains based on the properties of the $n$-magnon excitations
of the ferromagnetic vacuum, which has been conjectured by
Haldane [\ref{Haldane F D 1982 {\it J. Phys. C: Solid State Phys.}  {\bf 15}
L831; L1309}\refname\Halt].
In spin-$s$ chains the bound-state $n$-magnon dispersion branches
extend over
$\min (N,2s)$ Brillouin zones and integrability
manifests itself in that all branches are real
(i.e. lie outside the spin wave continuum)
and are continuous
through the zone-boundaries.
This integrability criterion has been  investigated for
2-magnon excitations in spin-$s$ chains in
[\ref{Chubukov A V and Kveschenko D V 1987
{\it J. Phys. C: Solid State Phys.} {\bf 20} L509}\refname\Chu,
\ref{Liu J M and Bonner J
C 1988 {\it J. App. Phys.} {\bf 63} 4114}\refname\Liu].
In particular, in [\Chu] the Haldane criterion was used to
obtain a relation
between the parameters of
a class of isotropic spin-$s$ hamiltonians.
This relation defines then a one parameter family of models,
supposed to be integrable.
But it contains only two out of the four known
fundamental integrable systems for $s=3/2$ [\Ken].
Neither the $su(4)$ invariant chain [\Suth] nor
the hamiltonian $H_{II}$ of [\BaYu] belong to it.
And apart from two points (describing the
Bethe-Ansatz integrable system
[\ref{Takhtajan L A 1982 {\it Phys. Lett.} {\bf 87A} 479; Babujian H M 1982
{\it Phys. Lett.} {\bf 90A} 479 }]
and the chain
related to the Temperley-Lieb algebra
[\ref{Perk J H H and Schultz C L 1983 {\it Physica} {\bf 122A} 50},
\ref{Affleck I 1990 {\it J. Phys. C: Cond. Mat.} {\bf 2} 405},
\ref{Barber M N  and Batchelor M T 1990 {\it J. Phys. A: Math. Gen}  {\bf 23}
L15}]), the other models
within the above one-parameter $s=3/2$ family are not fundamental
integrable systems, and are very likely to be nonintegrable.
Thus the condition in [\Chu] turns out to be
neither a necessary nor a sufficient condition for integrability. Presumably,
further
constraints should follow from the analysis of $n$-magnon
excitations with $n>2$. A better understanding of the Haldane
criterion,  as well as its relation to the
test proposed in this work is clearly needed.
Note that the former approach implies a choice of a
particular vacuum, while the $H_3$ test
is insensitive to the
ferromagnetic or antiferromagnetic character of the model.

\vfill\eject
\bigskip
\centerline{\sc{references}}
\immediate\closeout\refs \vskip 0.5cm
  \message{References}\input references

\end